\documentclass[11pt]{article} %Set document type and font size.

\usepackage{helvet}

\newcommand*{\myEXPfont}{\fontfamily{cmr}\selectfont}

\usepackage{graphicx} %Allows you to insert pdf and jpg images into your files.
\usepackage{mathtools}
\usepackage{amsmath}
\usepackage{mathrsfs}
\usepackage{multirow}
\usepackage[export]{adjustbox}
\usepackage[left=1in, right=1in, top=1in, bottom=1in]{geometry} %Sets the page margins as required.
\usepackage{booktabs}
\usepackage{rotating}
\usepackage{rotfloat}
\usepackage{float}
\usepackage{subfig}
\usepackage{url}
\usepackage{pdfpages}
\usepackage{rotating}
\usepackage[doublespacing]{setspace} %Double-space your text as required.
\usepackage{csvsimple}
\usepackage{array}
\usepackage{wrapfig}
\usepackage[utf8]{inputenc}

\usepackage[stable]{footmisc}

\usepackage{cite}
\usepackage{tabularx}
\usepackage{float}

\usepackage{amsthm}                % better theorem environments
\usepackage{amssymb}

\usepackage{multirow}

\usepackage{caption}
\usepackage{amsfonts}
\usepackage{amsmath} 
\usepackage{tikz}
\usetikzlibrary{positioning,shapes.geometric}

\usepackage{xcolor}

\usepackage{setspace}

\DeclareMathOperator{\E}{\mbox{{\myEXPfont E}}}

\makeatletter
\newcommand*{\indep}{%
  \mathbin{%
    \mathpalette{\@indep}{}%
  }%
}
\newcommand*{\nindep}{%
  \mathbin{%                   % The final symbol is a binary math operator
    \mathpalette{\@indep}{\not}% \mathpalette helps for the adaptation
  }%
}
\newcommand*{\@indep}[2]{%
  \sbox0{$#1\perp\m@th$}%        box 0 contains \perp symbol
  \sbox2{$#1=$}%                 box 2 for the height of =
  \sbox4{$#1\vcenter{}$}%        box 4 for the height of the math axis
  \rlap{\copy0}%                 first \perp
  \dimen@=\dimexpr\ht2-\ht4-.2pt\relax
  \kern\dimen@
  {#2}%
  \kern\dimen@
  \copy0 %                       second \perp
} 
\makeatother

\usepackage{xr-hyper}

\definecolor{forestgreen}{RGB}{34,139,34}
\usepackage{hyperref}
\hypersetup{
    colorlinks=true,
    linkcolor=magenta,
    filecolor=magenta, 
    citecolor=forestgreen,      
    urlcolor=blue
}

\usepackage{sectsty}
\sectionfont{\large}

\usepackage{layout}

\usepackage{geometry}

\newcolumntype{C}[1]{>{\centering\arraybackslash}p{#1}}

\usepackage{authblk}

\usepackage{afterpage}
\usepackage{lscape}
\usepackage{graphicx}

\usepackage{datetime}

\makeatletter
\newcommand*{\addFileDependency}[1]{% argument=file name and extension
  \typeout{(#1)}
  \@addtofilelist{#1}
  \IfFileExists{#1}{}{\typeout{No file #1.}}
}
\makeatother

\newcommand*{\myexternaldocument}[1]{%
    \externaldocument{#1}%
    \addFileDependency{#1.tex}%
    \addFileDependency{#1.aux}%
}

\myexternaldocument{appendix}

\begin{document}

\title{Estimating subgroup effects in generalizability and transportability analyses}

%\author[]{}

\author[1,2]{Sarah E. Robertson\thanks{Address for correspondence: Sarah E. Robertson; Box G-S121-8; Brown University School of Public Health, Providence, RI 02912; email: \texttt{sarah\_robertson@brown.edu}; phone: (401) 863-1000.}}
\author[3]{Jon A. Steingrimsson}
\author[4]{Nina R. Joyce}
\author[5]{Elizabeth A. Stuart}
\author[2,6]{Issa J. Dahabreh}

\affil[1]{Department of Health Services, Policy \& Practice, Brown University School of Public Health, Providence, RI}
\affil[2]{CAUSALab, Harvard T.H. Chan School of Public Health, Boston, MA}
\affil[3]{Department of Biostatistics, Brown University School of Public Health, Providence, RI }
\affil[4]{Department of Epidemiology, Brown University School of Public Health, Providence, RI}
\affil[5]{Departments of Mental Health, Biostatistics, and Health Policy and Management, Johns Hopkins Bloomberg School of Public Health, Baltimore, MD}
\affil[6]{Departments of Epidemiology and Biostatistics, Harvard T.H. Chan School of Public Health, Boston, MA}

\maketitle{}
\thispagestyle{empty} %cover page not numbered

\newpage
    \pagenumbering{arabic}
    \setcounter{page}{1}

%title page
\vspace*{0.2in}
{\bf \centering \large Estimating subgroup effects in generalizability and transportability analyses \par }

\iffalse

\vspace{0.2in}
\centerline{Sarah E. Robertson and Issa J. Dahabreh}

\vspace{0.5in}
\noindent Correspondence to Sarah E. Robertson, Box G-S121-8, Brown University, Providence, RI 02912 (e-mail: \texttt{sarah\_robertson@brown.edu})

\vspace{0.5in}
\noindent Author affiliations: Center for Evidence Synthesis in Health and Department of Health Services, Policy \& Practice, School of Public Health, Brown University, Providence, RI (Sarah E. Robertson); and Departments of Epidemiology and Biostatistics, Harvard T.H. Chan School of Public Health, Boston, MA (Issa J. Dahabreh).

\vspace{0.5in}
\noindent This work was funded by Patient-Centered Outcomes Research Institute (PCORI) award ME-1502-27794.

\fi

\vspace{0.5in}
\noindent \textbf{Conflict of interest:} None.

\vspace{0.5in}
\noindent \textbf{Running head:} Subgroup effects in generalizability and transportability analyses.

\vspace{0.5in}
\noindent
\textbf{Word count:} abstract= 156; main text $\approx$ 4000.

\vspace{0.5in}
\noindent
\textbf{Abbreviations:} \\
CASS = Coronary Artery Surgery Study \\
GAM = generalized additive model

%%%%%%%%%%%%%%%%%%%%%% ABSTRACT  %%%%%%%%%% %%%%%%%%%%%%%%%%
\clearpage
\vspace*{0.75in}
\begin{abstract}
\noindent {
Methods for extending -- generalizing or transporting -- inferences from a randomized trial to a target population involve conditioning on a large set of covariates that is sufficient for rendering the randomized and non-randomized groups exchangeable. Yet, decision-makers are often interested in examining treatment effects in subgroups of the target population defined in terms of only a few discrete covariates. Here, we propose methods for estimating subgroup-specific potential outcome means and average treatment effects in generalizability and transportability analyses, using outcome model-based (g-formula), weighting, and augmented weighting estimators. We consider estimating subgroup-specific average treatment effects in the target population and its non-randomized subset, and provide methods that are appropriate both for nested and non-nested trial designs. As an illustration, we apply the methods to data from the Coronary Artery Surgery Study to compare the effect of surgery plus medical therapy versus medical therapy alone for chronic coronary artery disease in subgroups defined by history of myocardial infarction.}

\vspace{0.3in}
\end{abstract}

%%%%%%%%%%%%%%%%%%%%%%%%%%%%%%%%%%%%%%%%%%%%%%%%%%%%%%%

%%%%%%%%%%%%%%%%%%%%%%%%%%%%%%%%%%%%%%%%%%%%%%%%%%%%%%%%%%%%%%%%%%%%%%%%%%%%%%%%%%%%%%%%%%%%%%%%%%%%%%%%%%
\clearpage
\section*{INTRODUCTION}
%%%%%%%%%%%%%%%%%%%%%%%%%%%%%%%%%%%%%%%%%%%%%%%%%%%%%%%%%%%%%%%%%%%%%%%%%%%%%%%%%%%%%%%%%%%%%%%%%%%%%%%%%%
When treatment effect modifiers have a different distribution among participants in a randomized trial compared to the target population of substantive interest, the average treatment effect estimated in the trial is not directly applicable to the target population. Methods for extending -- generalizing or transporting \cite{dahabreh2019commentaryonweiss} -- causal inferences from a trial to a target population of interest \cite{cole2010, westreich2017, dahabreh2018generalizing, dahabreh2020transportingStatMed, rudolph2017} rely on specifying models for either the probability of participation in the trial, the expectation of the outcome among trial participants, or both, to adjust for a set of covariates (typically measured at baseline) that is adequate to ensure conditional exchangeability between the randomized and non-randomized groups. Most work on extending causal inferences from trials has focused on methods for estimating average treatment effects in the entire target population or its non-randomized subset. Thus, analyses extending inferences from trials are characterized by the tension between the reason the methods are necessary (heterogeneous treatment effects over multiple covariates that have a different distribution between the randomized and non-randomized groups) and the output the methods produce (estimates of population-averaged treatment effects). 

Here, we propose methods for estimating subgroup-specific effects that can be used to alleviate this tension, when one or a few key discrete effect modifiers of substantive interest can be specified. Subgroup-specific average treatment effects in the trial may not generalize to the target population when there is a difference in the distribution of effect modifiers between the randomized and non-randomized groups, after conditioning on the subgroup variables \cite{seamans2021generalizability}. For such cases, the methods we describe can account for multiple variables that have a different distribution between randomized and non-randomized groups, while producing inferences appropriate for subgroups defined in terms of the smaller set of key effect modifiers.

Recently, Mehrotra et al. \cite{mehrotra2021transporting} described a weighting method for estimating subgroup-specific average treatment effects in the non-randomized subset of the target population. Their method can estimate subgroup-specific treatment effects only in the non-randomized subset of the target population. Furthermore, it relies on a model for the probability of trial participation but cannot be used for inference when estimating that model using data-adaptive methods (e.g., machine learning) \cite{chernozhukov2018double}. We build upon their work \cite{mehrotra2021transporting}, previous work on heterogeneity of treatment effects in observational studies over discrete subgroups \cite{eeren2015, xie_estimating_2012, robertson2020assessing}, and previous work on extending inferences from trials \cite{dahabreh2018generalizing, dahabreh2020transportingStatMed}, to describe estimators for subgroup-specific potential outcome means and average treatment effects, for the entire target population and its non-randomized subset. The estimators for the non-randomized subset of the target population can be used when the trial is nested in a sample of the target population (i.e., nested trial designs) or when the trial is combined with a separately obtained sample of non-randomized individuals (i.e., non-nested trial designs) \cite{dahabreh2021studydesigns}. We describe estimators appropriate for both study designs, including g-formula estimators that rely on models of the outcome among trial participants, weighting estimators that rely on models for the probability of trial participation, and augmented weighting estimators that combine both types of models. Our augmented weighting estimators are model doubly robust, in the sense that they are consistent when either the model for participation or the model for the outcome is correctly specified, but not necessarily both. Furthermore, they allow for the use of data-adaptive estimation methods. We illustrate the methods using data from the Coronary Artery Surgery Study (CASS) \cite{olschewski1985} to compare the effect of surgery plus medical therapy versus medical therapy alone for chronic coronary artery disease in subgroups defined by history of myocardial infarction.

%%%%%%%%%%%%%%%%%%%%%%%%%%%%%%%%%%%%%%%%%%%%%%%%%%%%%%%%%%%%%%%%%%%%%%%%%%%%%%%%%%%%%%%%%%%%%%%%%%%%%%%%%%
\section*{STUDY DESIGN, DATA, AND ESTIMANDS}
%%%%%%%%%%%%%%%%%%%%%%%%%%%%%%%%%%%%%%%%%%%%%%%%%%%%%%%%%%%%%%%%%%%%%%%%%%%%%%%%%%%%%%%%%%%%%%%%%%%%%%%%%%

\paragraph{Study design:} Both nested or non-nested trial designs can be used to extend inferences from a trial to a target population \cite{dahabreh2021studydesigns}. In a nested trial design, the trial is embedded in a sample of the target population, by prospectively recruiting trial participants within a cohort (e.g., in comprehensive cohort studies \cite{olschewski1985,olschewski1992, schmoor1996}) or by data linkage methods (e.g., using similar approaches as when linking a trial to administrative data to extend follow-up duration \cite{fitzpatrick2018assessment}). In a non-nested trial design, a sample from the non-randomized subset of the target population is obtained separately from the trial sample. 

In the nested trial design, it is possible to identify and estimate subgroup-specific causal effects in the entire target population and its non-randomized subset; in the non-nested trial design it is only possible to identify and estimate subgroup-specific causal effects in the non-randomized subset of the target population \cite{dahabreh2021studydesigns}. In a nested trial design, investigators may be interested both in estimating subgroups-specific effects in the entire target population or its non-randomized subset. Which estimand should be preferred in a nested trial design depends on the investigators' goals. For example, to inform clinical or policy decision-making, the effect in the entire target population is typically the most relevant estimand. In contrast, to compare treatment effects among the randomized and non-randomized groups, the effect in the non-randomized subset of the target population allows for a cleaner comparison against the effect in the trial.

In the remainder of the paper we work under the nested trial design because it is the design used in CASS. Nevertheless, our results for identifying and estimating subgroup-specific causal effects in the non-randomized subset of the target population can be extended to non-nested trial designs (see Web Appendix 1 for details) \cite{dahabreh2020transportingStatMed}.

\paragraph{Data:} Let $Y$ be an outcome measured at the end of the study (binary, continuous, or count); $A$ the assigned treatment strategy that takes values in a finite set $\mathcal A$; and $X$ a vector of covariates that takes values in $\mathcal{X}$. These covariates are typically measured before the decision to participate in the trial and before treatment assignment \cite{dahabreh2019identification} (or at least are not affected by trial participation and treatment assignment); we will refer to them as ``baseline covariates.'' Let $v$ be a proper subset of $\mathcal X$, $v \subset \mathcal X$. Then, $v$ defines a population subgroup consisting of individuals with $X \in v$. For example, we can define a collection of $k$ mutually exclusive and exhaustive non-empty subsets of $\mathcal{X}$, say $\{ v_1, \ldots, v_k \}$, that stratify the population into $k$ subgroups \cite{yang2021causal}. 

We are interested in examining heterogeneity over subgroups defined in terms of only a few discrete (or discretized continuous) covariates, when conditioning on a large set of covariates is necessary to ensure exchangeability between randomized and non-randomized groups (in the sense formalized in the next section). For instance, in our CASS analysis, we define subgroups based on the presence (vs. absence) of a myocardial infarction in an individual's medical history.  

The data collected in a nested trial design are realizations of the random tuple $(X_i, S_i, S_i A_i, S_i Y_i)$, with $i=1, \ldots, n$, where $n$ is the total sample size. In Web Appendix 1, we provide additional details regarding the data structure for non-nested trial designs.

\paragraph{Causal estimands:} Let $Y^{a}$ be the potential outcome under intervention to set treatment to $a$, for $a \in \mathcal A$ \cite{rubin1974, robins2000d}. When comparing two treatments, $a$ and $a^\prime$, both in $\mathcal A$, the subgroup-specific average treatment effect for subgroup $ X \in v$ in the target population is $\E[Y^{a} - Y^{a^\prime}| X \in v] = \E[Y^{a}| X \in v] - \E[Y^{a^\prime}| X \in v]$, and the subgroup-specific average treatment effect for subgroup $ X \in v$ in the non-randomized subset of the target population is $\E[Y^{a} - Y^{a^\prime}| X \in v, S = 0] = \E[Y^{a}| X \in v, S = 0] - \E[Y^{a^\prime}| X \in v, S = 0]$. These types of estimands are sometimes referred to as group-average treatment effects \cite{lechner2018modified}. Typically, in addition to these subgroup-specific average treatment effects, we are also interested in their component potential outcome means, $\E[Y^{a}| X \in v]$ and $\E[Y^{a}| X \in v, S=0]$, for each $a \in \mathcal A$.

%%%%%%%%%%%%%%%%%%%%%%%%%%%%%%%%%%%%%%%%%%%%%%%%%%%%%%%%%%%%%%%%%%%%%%%%%%%%%%%%%%%%%%%%%%%%%%%%%%%%%%%%%%
\section*{IDENTIFICATION}
%%%%%%%%%%%%%%%%%%%%%%%%%%%%%%%%%%%%%%%%%%%%%%%%%%%%%%%%%%%%%%%%%%%%%%%%%%%%%%%%%%%%%%%%%%%%%%%%%%%%%%%%%%

\subsection*{Identification of potential outcome means in the entire target population}

\paragraph{Identifiability conditions:} 
We now list sufficient conditions under which the subgroup-specific potential outcome means and subgroup-specific average treatment effect in the target population are identifiable. These same conditions are also sufficient for identifying the overall potential outcome means and average treatment effects in the target population \cite{dahabreh2019generalizing, dahabreh2019identification}. 

%Suppose that the following conditions hold  \cite{dahabreh2018generalizing,dahabreh2019identification}: 

\noindent
\emph{(1) Consistency of potential outcomes:} for each unit $i$ in the target population and each $a \in \mathcal A$, if $A_i = a$, then $Y_i = Y^{a}_i$.

\noindent
\emph{(2) Mean exchangeability in the trial over $A$:} for every $a \in \mathcal A$ and every $x$ with positive density in the trial $f(x, S = 1 ) \neq 0$, $\E [ Y^{a} | X = x , S = 1, A =a] = \E [ Y^{a} | X = x, S = 1]$.

\noindent
\emph{(3) Positivity of treatment assignment:} $\Pr[A=a | X = x, S=1] > 0$, for each $a \in \mathcal A$ and each $x$ with positive density in the trial $f(x , S = 1) \neq 0$.

\noindent
\emph{(4) Mean exchangeability over $S$:} $\E [ Y^a | X = x , S = 1] = \E [ Y^a | X = x ] $, for every $x$ with positive density in the target population $f(x) \neq 0$ and for each $a \in \mathcal A$.

\noindent
\emph{(5) Positivity of trial participation:} $\Pr[S=1 | X = x] >0,$ for every $x$ such that $f(x) \neq 0.$ 

Note that if interest focuses on a single subgroup, then the positivity and exchangeability assumptions can be weakened to apply only to covariate values that can occur in the specific subgroup. In most applications, however, we are interested in some collection of subgroups that are mutually exclusive and exhaustive of $\mathcal X$, not just a single subgroup.

In addition, to focus on estimation issues, we consider the above identifiability conditions without reference to a fully elaborated structural model. Nevertheless, the conditions we invoke can be derived from structural equation models for the data generating mechanism, such as causal directed acyclic graphs and single-world intervention graphs \cite{richardson2013single, dahabreh2019identification, dahabreh2020benchmarking} or selection diagrams \cite{pearl2014}.

%In particular, it has been recently discussed  \cite{dahabreh2019identification, dahabreh2020benchmarking} how the distributional exchangeability conditions can be derived from non-parametric structural equation models with a finest fully randomized causally interpretable structured tree graph error structure, represented using causal directed acyclic graphs and single-world intervention graphs \cite{pearl2014, richardson2013single} (also see references for an alternative approach based on selection diagrams).

\paragraph{Reasoning about the identifiability conditions:} In applications, the identifiability conditions need to be examined in light of background knowledge and possibly subjected to sensitivity analyses \cite{dahabreh2019commentaryonweiss, dahabreh2019sensitivity}. Condition 1 requires the absence of ``hidden'' versions of treatment (or at least treatment variation irrelevance among versions) \cite{rubin1986, rubin2010reflections, vanderWeele2009}, trial engagement effects \cite{dahabreh2019identification, dahabreh2019generalizing}, and interference among individuals \cite{rubin1986, halloran1995causal}. This condition is largely untestable and judgements about its plausibility should be informed by substantive knowledge about the trial, the sampling of the target population, and the treatments of interest. Conditions 2 and 3 are likely to hold by design in marginally or conditionally randomized trials \cite{hernan2020causal}. Condition 4 is also untestable and judgements about its plausibility should be informed by substantive knowledge and sensitivity analyses \cite{dahabreh2019sensitivity}. Condition 5 is in principle testable, but can be challenging to assess when $X$ is high-dimensional \cite{petersen2012diagnosing}.

Furthermore, in applications, other complications, such as missing data, loss to follow-up, or non-adherence, may require additional conditions for identification. To focus on issues related to extending inferences from trials, we work in the setting with complete data, no losses to follow-up, and complete adherence. Nevertheless, the methods we describe can be naturally extended to address these complications \cite{dahabreh2019identification}.

\paragraph{Identification:} As we show in Web Appendix 2, under conditions 1 through 5, the subgroup-specific potential outcome mean in the target population with $X \in v$, $\E[Y^{a}| X \in v]$, for each $a \in \mathcal A$, is identified by 
\begin{equation}\label{id_g_formula}
    \begin{split}
   \psi(v, a) \equiv \E\big[\E[Y | X, S  = 1, A = a] \big| X \in v \big].
   \end{split}
\end{equation}
Under the positivity conditions, $\psi(v, a)$ has an algebraically equivalent inverse probability weighting \cite{robins2000a, dahabreh2019relation} re-expression,
\begin{equation} \label{id_IP_weighting1}
    \psi(v, a) =  \dfrac{1}{\Pr[X \in v]} \E\left[ \dfrac{ I(S = 1, A = a, X \in v)Y}{\Pr[S = 1 | X] \Pr[A = a | X, S = 1]}  \right],
\end{equation}
where $I(\cdot)$ denotes the indicator function, and the terms $\dfrac{ I(S = 1, A = a)}{\Pr[S = 1 | X] \Pr[A = a | X, S = 1]}$ inside the expectation can be viewed as population inverse probability of participation and treatment weights. Under positivity, we can replace the denominator in the first term of equation \eqref{id_IP_weighting1} with the expectation of these weights, because
\begin{equation*}
   \Pr[X \in v] = \E\left[\dfrac{I(S = 1, A = a, X \in v)}{\Pr[S = 1 | X] \Pr[A = a | X, S = 1]}\right].
\end{equation*}

\subsection*{Identification of potential outcome means in the non-randomized subset}

\paragraph{Identifiability conditions:} To identify the potential outcome means in the non-randomized subset, we retain conditions 1 through 3 as listed above, but substitute the following for conditions 4 and 5 \cite{dahabreh2020transportingStatMed, dahabreh2019identification}:

\noindent
\emph{(4\textsuperscript{*}) Mean exchangeability over $S$:} $\E [ Y^a | X = x , S = 1] = \E [ Y^a| X = x, S=0 ] $, for every $x$ with positive density in the non-randomized subset of the target population $f(x, S = 0) \neq 0$ and for each $a \in \mathcal A$. 

\noindent
\emph{(5\textsuperscript{*}) Positivity of trial participation:} $\Pr[S=1 | X = x] >0,$ for every $x$ such that $f(x, S=0) \neq 0.$

Conditions 1 through 3, along with 4\textsuperscript{*} and 5\textsuperscript{*}, are sufficient for identifying the subgroup-specific potential outcome means and average treatment effects in the non-randomized subset of the target population. The same conditions are also sufficient for identifying the overall potential outcome means and average treatment effects in the non-randomized subset of the target population \cite{dahabreh2020transportingStatMed, dahabreh2019identification}. Reasoning about conditions 4\textsuperscript{*} and 5\textsuperscript{*} is similar to that for conditions 4 and 5. 

%\vspace{0.2cm}
%\noindent 
\paragraph{Identification:} As we show in Web Appendix 2, under conditions 1 through 3, 4\textsuperscript{*}, and 5\textsuperscript{*}, the subgroup-specific potential outcome mean in the non-randomized subset of the target population with $X \in v$, $\E[Y^{a}| X \in v, S = 0]$, for each $a \in \mathcal A$, is identified by 
\begin{equation}\label{id_g_formula_S0}
    \begin{split}
   \phi(v, a) \equiv \E\big[\E[Y | X, S  = 1, A = a] \big| X \in v , S = 0 \big].
   \end{split}
\end{equation}
Under the positivity conditions, $\phi(v, a)$ has an algebraically equivalent inverse odds weighting \cite{robins2000a, dahabreh2019relation} re-expression,
\begin{equation} \label{id_IP_weighting1_S0}
    \phi(v, a) =  \dfrac{1}{\Pr[X \in v, S=0]} \E\left[ \dfrac{ I(S = 1, A = a, X \in v)\Pr[S=0|X]Y}{\Pr[S = 1 | X] \Pr[A = a | X, S = 1]}  \right].
\end{equation}
Here, the terms $\dfrac{ I(S = 1, A = a)\Pr[S=0|X]}{\Pr[S = 1 | X] \Pr[A = a | X, S = 1]}$ inside the expectation can be viewed as population inverse odds of participation and inverse probability of treatment weights. Under positivity, we can replace the denominator in the first term of equation \eqref{id_IP_weighting1_S0}, with the expectation of these weights, because
\begin{equation*} \label{id_IP_weighting2_S0} 
    \Pr[X \in v, S=0] = \E\left[\dfrac{I(S = 1, A = a, X \in v)\Pr[S=0|X]}{\Pr[S = 1 | X] \Pr[A = a | X, S = 1]}\right].
\end{equation*}

\subsection*{Identification of treatment effects}
The previous results identify the subgroup-specific potential outcomes means in the target population or its non-randomized subset. It follows that we can identify subgroup-specific average treatment effects by taking differences and that we can identify measures of heterogeneity as contrasts of these subgroup-specific effects. For example, comparing treatments $a$ and $a^\prime$, under conditions 1 through 5, the subgroup-specific average treatment effect in the target population is identified by $\psi(v, a) - \psi(v,a^\prime)$. Similarly, under conditions 1 through 3, 4\textsuperscript{*}, and 5\textsuperscript{*}, the subgroup-specific average treatment effect in the non-randomized subset of the target population is identified by $\phi(v, a) - \phi(v,a^\prime)$.

%%%%%%%%%%%%%%%%%%%%%%%%%%%%%%%%%%%%%%%%%%%%%%%%%%%%%%%%%%%%%%%%%%%%%%%%%%%%%%%%%%%%%%%%%%%%%%%%%%%%%%%%%%
%\section{Estimating subgroup-specific potential outcome means}
\section*{ESTIMATION \& INFERENCE}
%%%%%%%%%%%%%%%%%%%%%%%%%%%%%%%%%%%%%%%%%%%%%%%%%%%%%%%%%%%%%%%%%%%%%%%%%%%%%%%%%%%%%%%%%%%%%%%%%%%%%%%%%%

\subsection*{Estimation of potential outcome means in the entire target population}

\paragraph{Outcome modeling and standardization (g-formula):} The identification result in equation \eqref{id_g_formula} suggests the following outcome model-based estimator of $\psi(v,a)$,
\begin{equation} \label{g_formula_estimator}
	\widehat {\psi}_{\text{\tiny{OM}}}(v,a) = \left\{ \sum_{i=1}^{n} I(X_i \in v) \right\}^{-1} \sum\limits_{i = 1}^{n} I(X_i \in v) \widehat g_{a}(X_i),
\end{equation}
where $\widehat g_{a}(X)$ is an estimator for $\E [Y | X, S=1, A = a]$. Typically, $\widehat g_{a}(X)$ is obtained using a parametric model for the conditional expectation of the outcome given treatment $A = a$ and the baseline covariates. If the parametric model is correctly specified, $\widehat g_{a}(X)$ is a consistent estimator for $\E [Y | X, S=1, A = a]$ and $\widehat {\psi}_{\text{\tiny{OM}}}(v,a)$ is a consistent estimator for $\psi(v,a)$ \cite{hernan2020causal, dahabreh2019relation}. 

\paragraph{Inverse probability weighting:} The identification result in equation \eqref{id_IP_weighting1} suggests an inverse probability weighted estimator: 
\begin{equation} \label{IPW_estimator1}
  \widehat {\psi}_{\text{\tiny{IPW1}}}(v,a) = \left\{ \sum_{i=1}^{n}  I(X_i \in v) \right\}^{-1} \sum\limits_{i = 1}^{n} \widehat w_{v,a}(X_i, S_i, A_i) Y_i,
\end{equation}
where the weight for the $i$th observation is defined as $$\widehat w_{v,a}(X_i, S_i, A_i) = \dfrac{I(S_i = 1, A_i = a,X_i \in v)}{\widehat p(X_i)\widehat e_a(X_i)},$$ $\widehat p(X)$ is an estimator for $\Pr[S=1|X],$ and $\widehat e_a(X)$ is an estimator for $\Pr[ A = a | X , S=1]$. 
We can obtain another inverse probability weighting estimator by normalizing the weights \cite{hajek1971comment}:
\begin{equation} \label{IPW_estimator2}
  \widehat {\psi}_{\text{\tiny{IPW2}}}(v,a) = \left\{ \sum_{i=1}^{n} \widehat w_{v,a}(X_i, S_i, A_i)  \right\}^{-1} \sum\limits_{i = 1}^{n} \widehat w_{v,a}(X_i, S_i, A_i)  Y_i.
\end{equation}
Typically, $\widehat p(X)$ is obtained using a parametric model for the probability of participation, conditional on baseline covariates. Similarly, $\widehat e_{a}(X)$ is often obtained using a parametric model for the probability of treatment. Typically, the probability of treatment is known, so $\widehat e_{a}(X)$ cannot be misspecified, and the known probability can be used in place of $\widehat e_{a}(X)$; but estimating $\widehat e_{a}(X)$ may correct for slight imbalances in the trial and improve efficiency \cite{lunceford2004, williamson2014variance}. Thus, in most trials, if the participation model is correctly specified, $\widehat p(X)$ is a consistent estimator for $\Pr[S=1|X]$, and both $\widehat {\psi}_{\text{\tiny{IPW1}}}(v,a)$ and $\widehat {\psi}_{\text{\tiny{IPW2}}}(v,a)$ are consistent estimators for $\psi(v,a)$ \cite{hernan2020causal, dahabreh2019relation}. In Web Appendix 3, we describe how an estimator equivalent to $\widehat {\psi}_{\text{\tiny{IPW2}}}(v,a)$ can be obtained by fitting a weighted saturated regression model for the outcome conditional on treatment and subgroup indicators.

\paragraph{Augmented inverse probability weighting:} The estimators above require models for either the expectation of the outcome or the probability of participation; we can obtain an estimator that uses both models: 
\begin{align} \label{DR_estimator1}
\begin{split}
 \widehat {\psi}_{\text{\tiny{AIPW1}}}(v,a) &\stackrel{\text{}}= \left\{ \sum_{i=1}^{n} I(X_i \in v) \right\}^{-1} \sum_{i=1}^{n}  \Big\{ \widehat w_{v,a}(X_i, S_i, A_i) \left\{Y_i- \widehat g_{a}(X_i)\right\}  +  I(X_i \in v) \widehat g_{a}(X_i) \Big\},
\end{split}
\end{align}
where $\widehat w_{v,a}(X, S, A)$ and $\widehat g_a(X)$ are as defined above. We can also obtain a second augmented inverse probability weighting estimator by normalizing the weights:
\begin{align} \label{DR_estimator2}
\begin{split}
 \widehat {\psi}_{\text{\tiny{AIPW2}}}(v,a) &\stackrel{\text{}}= \left\{ \sum_{i=1}^{n} \widehat w_{v,a}(X_i, S_i, A_i) \right\}^{-1} \sum_{i=1}^{n}  \widehat w_{v,a}(X_i, S_i, A_i) \left\{Y_i- \widehat g_{a}(X_i)\right\}  \\
&\quad\quad\quad + \left\{ \sum_{i=1}^{n} I(X_i \in v) \right\}^{-1} \sum_{i=1}^{n}  I(X_i \in v) \widehat g_{a}(X_i).
\end{split}
\end{align}
When either $\widehat p(X)$ or $\widehat g_a(X)$ is obtained from a correctly specified model, $\widehat {\psi}_{\text{\tiny{AIPW1}}}$ and $\widehat {\psi}_{\text{\tiny{AIPW2}}}$ are consistent estimators for $\psi(v,a)$. Thus, $\widehat {\psi}_{\text{\tiny{AIPW1}}}$ and $\widehat {\psi}_{\text{\tiny{AIPW2}}}$ are model doubly robust, in the sense that they are consistent when either the model for the expectation of the outcome or the model for the probability of participation is correctly specified, but not necessarily both \cite{bang2005}. They are also rate doubly robust \cite{smucler2019unifying}: if data-adaptive methods are used to estimate $\widehat p(X)$ or $\widehat g_a(X)$ with a fast-enough rate \cite{chernozhukov2018double} (but slower than parametric), the estimators still converge at $\sqrt{n}$-rate. For completeness, in Web Appendix 4 we describe another estimator that relies on fitting a weighted multivariable regression model for the outcome, followed by standardization over the distribution of covariates in the target population \cite{robins2007, wooldridge2007}.

%When either $\widehat p(X)$ or $\widehat g_a(X)$ is correctly specified, $\widehat {\psi}_{\text{\tiny{AIPW1}}}$ is doubly robust  \cite{bang2005} and is a consistent estimator for $\psi(v,a)$.

%The augmented inverse probability  weighting estimator is doubly robust \cite{bang2005}, in the sense that it is consistent when at least one of the models it relies on is correctly specified. It converges in probability to $\psi(v,a)$ when either the model for the probability of participation or the model for the expectation of the outcome is correctly specified . 

\subsection*{Estimation of potential outcome means in the non-randomized subset}

\paragraph{Outcome modeling and standardization (g-formula):} The identification result in equation \eqref{id_g_formula_S0} suggests the following outcome model-based estimator for $\phi(v, a)$:
\begin{equation} \label{g_formula_estimator_S0}
	\widehat {\phi}_{\text{\tiny{OM}}}(v,a) = \left\{ \sum_{i=1}^{n} I(X_i \in v,S_i=0) \right\}^{-1} \sum\limits_{i = 1}^{n} I(X_i \in v,S_i=0) \widehat g_{a}(X_i).
\end{equation}
If the parametric model for the expectation of the outcome is correctly specified, $\widehat g_{a}(X)$ is a consistent estimator for $\E [Y | X, S=1, A = a]$ and $\widehat {\phi}_{\text{\tiny{OM}}}(v,a)$ is a consistent estimator for $\phi(v,a)$ \cite{hernan2020causal, dahabreh2019relation}. 

%When a parametric outcome model $g_{a}(X)$ is correctly specified, the estimator $\widehat {\phi}_{\text{\tiny{OM}}}(v,a)$ converges in probability to $\phi(v,a)$ \cite{hernan2020}.

\paragraph{Inverse odds weighting:} The identification result in \eqref{id_IP_weighting1_S0} suggests an inverse odds weighting estimator: 
\begin{equation} \label{IPW_estimator1_S0}
  \widehat {\phi}_{\text{\tiny{IOW1}}}(v,a) = \left\{ \sum_{i=1}^{n}  I(X_i \in v,S_i=0) \right\}^{-1} \sum\limits_{i = 1}^{n} \widehat o_{v,a}(X_i, S_i,A_i) Y_i,
\end{equation}
where the inverse odds weight for the $i$th observation is defined as $$\widehat o_{v,a}(X_i,S_i,A_i) =\dfrac{I(S_i = 1, A_i = a,X_i \in v) \big\{ 1-\widehat p(X_i) \big\}}{\widehat p(X_i)\widehat e_a(X_i)}.$$ Of note, we refer to these weights as ``inverse odds weights'' because their defining characteristic is the term for the inverse of the odds of trial participation, $\{ 1-\widehat p(X_i) \}/\widehat p(X_i)$, even though they also contain a component for the inverse of the estimated probability of receiving the treatment actually received (in the trial), $1/\widehat e_a(X_i)$. 

We can obtain another inverse odds weighting estimator by normalizing the weights \cite{hajek1971comment}: 
\begin{equation} \label{IPW_estimator2_S0}
  \widehat {\phi}_{\text{\tiny{IOW2}}}(v,a) = \left\{ \sum_{i=1}^{n}  \widehat o_{v,a}(X_i, S_i,A_i) \right\}^{-1} \sum\limits_{i = 1}^{n}  \widehat o_{v,a}(X_i, S_i,A_i) Y_i.
\end{equation}
If the parametric model for the probability of participation is correctly specified, $\widehat p(X)$ is a consistent estimator for $\Pr[S=1|X]$ and both $\widehat {\phi}_{\text{\tiny{IOW1}}}(v,a)$ and $\widehat {\phi}_{\text{\tiny{IOW2}}}(v,a)$ are consistent estimators for $\phi(v,a)$ \cite{hernan2020causal}. In Web Appendix 3 we describe how an estimator equivalent to $\widehat {\phi}_{\text{\tiny{IOW2}}}(v,a)$ can be obtained by fitting an inverse odds weighted saturated regression model for the outcome conditional on treatment and subgroup indicators; this estimator relates but is not identical to the one described in reference \cite{mehrotra2021transporting} (see Web Appendix 3 for details).

%\vspace{0.3cm}
%\noindent 
\paragraph{Augmented inverse odds weighting:} One augmented inverse odds weighting estimator of the subgroup-specific potential outcome mean is:
\begin{align} \label{DR_estimator1_S0}
\begin{split}
 \widehat {\phi}_{\text{\tiny{AIOW1}}}(v,a) &\stackrel{\text{}}= \left\{ \sum_{i=1}^{n} I(X_i \in v, S_i=0) \right\}^{-1} \sum_{i=1}^{n}  \Big\{    \widehat o_{v,a}(X_i, S_i,A_i) \big\{ Y_i- \widehat g_{a}(X_i) \big\}  \\
&\quad\quad\quad\quad\quad\quad\quad\quad\quad\quad\quad\quad\quad\quad\quad + I(X_i \in v,S_i=0) \widehat g_{a}(X_i) \Big\}.
\end{split}
\end{align}
We can obtain a second augmented inverse odds weighting estimator by normalizing the weights:
\begin{align} \label{DR_estimator2_S0}
\begin{split}
 \widehat {\phi}_{\text{\tiny{AIOW2}}}(v,a) &\stackrel{\text{}}= \left\{ \sum_{i=1}^{n} \widehat o_{v,a}(X_i, S_i,A_i) \right\}^{-1} \sum_{i=1}^{n}  \widehat o_{v,a}(X_i, S_i,A_i) \left\{Y_i- \widehat g_{a}(X_i)\right\}  \\
&\quad\quad\quad + \left\{ \sum_{i=1}^{n} I(X_i \in v,S_i=0) \right\}^{-1} \sum_{i=1}^{n}  I(X_i \in v,S_i=0) \widehat g_{a}(X_i).
\end{split}
\end{align}
These augmented inverse odds weighting estimators, like the augmented inverse probability weighting estimators discussed above, are also model and rate doubly robust (for related results see \cite{dahabreh2019generalizing}). Last, in Web Appendix 4, we describe another estimator that relies on fitting an inverse odds weighted multivariable regression model for the outcome, followed by standardization over the distribution of baseline covariates in the non-randomized subset of the target population.

\subsection*{Estimation of subgroup-specific treatment effects}

Subgroup-specific average treatment effects can be estimated by taking differences between pairs of the potential outcome mean estimators described above. For example, the subgroup-specific average treatment effect in the entire target population, comparing treatments $a$ and $a^\prime$, using the augmented inverse probability weighting estimator in equation \eqref{DR_estimator1}, can be estimated as $\widehat \psi_{\text{\tiny AIPW1}}(v, a) - \widehat \psi_{\text{\tiny AIPW1}}(v,a^\prime)$. Analogous treatment effect estimators can be obtained using each of the potential outcome mean estimators in equations \eqref{g_formula_estimator} through \eqref{DR_estimator2}, for the entire target population, or those in equations \eqref{g_formula_estimator_S0} through \eqref{DR_estimator2_S0}, for the non-randomized subset of the target population.

\subsection*{Modeling to estimate subgroup-specific treatment effects}

\paragraph{Borrowing strength across subgroups:} It is important to note that the methods we describe do not require fitting separate models for the expectation of the outcome or the probability of trial participation in each subgroup. The methods do require correct specification of these models, which may require specifying product terms (statistical interactions) between the subgroup defining variables and other covariates, but not necessarily separate modeling in each subgroup. For example, consider fitting a parametric regression model for the probability of trial participation conditional on covariates, separately in each subgroup. Fitting this model separately in each subgroup is practically equivalent to using all possible subgroup-covariate product terms in the regression. But if some of the coefficients of these product terms are zero (or very close to zero), then we can borrow strength across subgroups by fitting one model for participation that omits the unnecessary product terms. This is useful when, as is commonly the case in applications, some subgroups in the trial are small, making the approach of fitting each model separately by subgroup infeasible \cite{senn2008statistical}.

Thus, our identification and estimation results show that splitting the data by the subgroup-defining variables is not needed for subgroup-specific generalizability or transportability analyses. Although splitting the trial and target population sample by the subgroup-defining variables and performing the analysis in each subgroup separately is a natural approach (previously proposed estimators for extending inferences \cite{cole2010, westreich2017, dahabreh2018generalizing, dahabreh2020transportingStatMed} can be used with no modifications to estimate subgroup-specific effects) such a strategy can be infeasible when the subgroup is a small part of the trial or the target population and will often be inefficient.

\paragraph*{Data-adaptive estimation:} Parametric models are the most common approach for estimating the probability of trial participation and the expectation of the outcome among trial participants. Because the probability of participation and the expectation of the outcome are unknown and must be estimated, and parametric models are likely misspecified, it may be desirable to use data-adaptive modeling approaches (e.g., machine learning techniques). When using such data-adaptive approaches, the augmented weighting estimators presented above can still support valid inference because they can accommodate rates of convergence slower than the parametric rate, when estimating the probability of trial participation and the expectation of the outcome among trial participants \cite{chernozhukov2018double}. For many data-adaptive approaches it may also be necessary to modify our estimators to use cross-fitting strategies, as described in prior work \cite{chernozhukov2018double, chernozhukov2017double}. Of note, estimating the probability of treatment among trial participants is typically not necessary because the known-by-design randomization probability can be used, but we recommend estimating it with a parametric model because that model cannot be misspecified and its use may improve estimate precision. 

\subsection*{Inference}

For all the estimators described above, inference using standard M-estimation methods \cite{stefanski2002} is possible (see references \cite{lunceford2004, williamson2014variance,yang2021causal} for examples). These methods can correctly account for the estimation of the working models for the probability of participation and the expectation of the outcome (and the probability of treatment, if needed) when obtaining measures of uncertainty. Furthermore, when using parametric models, we have found that numerical approximations to the usual M-estimation sandwich variance, such as those provided by the \texttt{geex} package \cite{saul2020calculus} in \texttt{R} \cite{currentRcitation}, work quite well in practice. Inference using non-parametric bootstrap methods \cite{efron1994introduction} is also easy to obtain and will often be preferred because of the simplicity of implementing resampling methods in most standard software packages.

%%%%%%%%%%%%%%%%%%%%%%%%%%%%%%%%%%%%%%%%%%%%%%%%%%%%%%%%
\section*{SUBGROUP ANALYSES IN CASS}
%%%%%%%%%%%%%%%%%%%%%%%%%%%%%%%%%%%%%%%%%%%%%%%%%%%%%%%%

%\vspace{0.3cm}
%\noindent
\paragraph{CASS design and data:} The Coronary Artery Surgery Study (CASS) was a comprehensive cohort study \cite{olschewski1985} that compared coronary artery bypass grafting surgery plus medical therapy (hereafter ``surgery'') versus medical therapy alone for patients with chronic coronary artery disease. The data from a comprehensive cohort study, such as CASS, are consistent with the nested trial study design \cite{dahabreh2021studydesigns}. In CASS, among a total of 2099 trial-eligible patients, 780 agreed to participate in the randomized study; 1319 declined and were enrolled into an observational study of the same treatments. We excluded six patients for consistency with prior CASS analyses \cite{chaitman1990, olschewski1992} and in accordance with CASS data release notes. CASS recruited patients from August 1975 to May 1979, with follow-up until December 1996. 

CASS collected a common set of baseline covariates using the same measurement approach among both the randomized and non-randomized groups. Additionally, the setting (in terms of doctors, location, and time) is similar between the randomized and non-randomized groups. In fact, the investigators who conducted CASS have, in prior work \cite{olschewski1992analysis}, jointly analyzed data from the randomized and non-randomized groups, suggesting that they would deem the consistency and exchangeability conditions (i.e., conditions 1 and 4, listed above) to be plausible \cite{dahabreh2020benchmarking}. Like most analyses attempting to draw causal inferences using data sources beyond a single randomized trial, our analyses of the CASS data invoke conditions that are not guaranteed to hold and are untestable. Nevertheless, we believe that the design of CASS makes the conditions reasonably plausible.

%\vspace{0.3cm}
%\noindent
\paragraph{Estimators:} We applied the estimators presented above to estimate the 10-year risk of death (cumulative incidence proportion) and risk difference in subgroups defined by history of myocardial infarction, for the entire target population and its non-randomized subset. Because no patients were lost to follow-up during the first ten years of the study, the cumulative incidence proportion of death is a reasonable measure of risk. We provide code to implement the methods in Web Appendix 6.

%\vspace{0.3cm}
%\noindent
\paragraph{Model specification:} We fit logistic regression models for the probability of trial participation, the probability of the outcome among individuals randomized in each treatment group, and the probability of treatment in the trial. All models adjusted for the main effects of covariates that we used in a previous analysis \cite{dahabreh2018generalizing} as well as an original CASS analysis that combined data from the randomized and observational components of the study \cite{olschewski1992}: age, severity of angina, history of previous myocardial infarction, percent obstruction of the proximal left anterior descending artery, left ventricular wall motion score, number of diseased vessels, and ejection fraction. For simplicity, we restricted our analyses to patients with complete data on the baseline covariates we adjusted for. In previous work \cite{dahabreh2018generalizing}, using weighting and multiple imputation methods under a missing-at-random assumption, we found that accounting for missing data did not appreciably affect estimates (and their uncertainty). To allow for more flexible modeling, we also used a generalized additive model (GAM) to estimate the probability of participation and the probability of the outcome. We used the \texttt{gam} function in the \texttt{mgcv} package (v1.8-34) \cite{Wood2011-ax, woodGAMbook} in \texttt{R} (v4.0.4) \cite{currentRcitation}, which fits a smoothness penalty using generalized cross-validation and we considered splines for age and ejection fraction.
%which fits penalized thin plate regression splines and allows up to 10 for the dimension of the basis. 

%\vspace{0.3cm}
%\noindent
\paragraph{Results:} We analyzed data from 1686 individuals; 731 randomized (368 to surgery and 363 to medical therapy) and 955 non-randomized. We summarize the baseline characteristics of individuals included in the analysis in Web Table 2 of Web Appendix 5. Estimates of the 10-year risk differences are shown for the target population of all trial-eligible patients (Table \ref{table_cass_analyses1}) and for its non-randomized subset (Table \ref{table_cass_analyses2}). We obtained non-parametric bootstrap-based standard errors and used them to calculate 95\% Wald-style confidence intervals for the 10-year mortality risk and risk difference (10,000 resampled datasets). We report results for the 10-year mortality risk for each treatment in each subgroup in Web Table 2. The results were similar across different estimators, suggesting that our modeling choices are reasonable \cite{robins2001}. Among patients without a history of previous myocardial infarction, we estimated a benefit of approximately 2\% in favor of medical therapy alone; among patients with a history of myocardial infarction we estimated a benefit of approximately 5\% in favor of surgery. The confidence intervals for both subgroups were wide and overlapping, reflecting considerable uncertainty in the subgroup analysis.

%%%%%%%%%%%%%%%%%%%%%%%%%%%%%%%%%%%%%%%%%%%%%%%%%%%%%%%%
\section*{DISCUSSION}
%%%%%%%%%%%%%%%%%%%%%%%%%%%%%%%%%%%%%%%%%%%%%%%%%%%%%%%%

We proposed g-formula, weighting, and augmented weighting estimators for subgroup-specific potential outcome means and average treatment effects in the entire target population and its non-randomized subset for nested trial designs. The estimators for the non-randomized subset of the target population are also applicable to non-nested trial designs, when the trial and non-randomized sample of observations are obtained separately \cite{dahabreh2021studydesigns}. Our augmented weighting estimators, which combine models for the probability of trial participation and the expectation of the outcome, may be preferred in practice because they offer robustness to model misspecification \cite{bang2005} and allow for valid inference when the models are estimated using data-adaptive methods \cite{chernozhukov2018double}, including machine learning techniques \cite{friedman2001elements, bishop2006pattern}. 

The methods we describe should be useful for confirmatory or descriptive subgroup analyses \cite{varadhan2013framework} that examine heterogeneity of treatment effects over subgroups defined in terms of one or more key effect modifiers. Key effect modifiers are often identified in trial protocols on the basis of substantive knowledge (e.g., when pre-specifying subgroup analyses) \cite{dahabreh2017bookchapter}. Subgroup-specific treatment effects conditional on these effect modifiers are also of interest when extending inferences from trials to target populations. Estimates of these subgroup-specific treatment effects can be useful in the presence of strong effect modification, when the average treatment effect is not sufficient for guiding decisions in the target population, particularly in limited resource settings when we need to focus intervention efforts on subgroups that are most likely to benefit from treatment \cite{rothwell2005subgroup}. Furthermore, the methods we describe can form the basis of exploratory approaches for data-driven subgroup discovery (e.g., by adapting tree-based subgroup discovery approaches \cite{yang2021causal} to generalizability and transportability analyses).

%%%%%%%%%%%%%%%%%%%%%%%%%%%%%%%%%%%%%%%%%%%%%%%%%%%%%%%%%%%%
\section*{ACKNOWLEDGEMENTS}
%%%%%%%%%%%%%%%%%%%%%%%%%%%%%%%%%%%%%%%%%%%%%%%%%%%%%%%%%%%%

This work was supported in part by Agency for Healthcare Research and Quality (AHRQ) award R36HS028373-01 and Patient-Centered Outcomes Research Institute (PCORI) award ME-1306-03758. The content is solely the responsibility of the authors and does not necessarily represent the official views of PCORI, its Board of Governors, the PCORI Methodology Committee, or AHRQ. The data analyses in our paper used CASS research materials obtained from the NHLBI Biologic Specimen and Data Repository Information Coordinating Center. This paper does not necessarily reflect the opinions or views of the CASS or the NHLBI. Conflict of interest: none declared.

%%%%%%%%%%%%%%%%%%%%%%%%%%%%%%%%%%%%%%%%%%%%%%%%%%%%%%%%%%%%
%%%%%%%%%%%%%%%%%%%%%%%%%REFERENCES%%%%%%%%%%%%%%%%%%%%%%%%%
%%%%%%%%%%%%%%%%%%%%%%%%%%%%%%%%%%%%%%%%%%%%%%%%%%%%%%%%%%%%

\clearpage
\bibliographystyle{ieeetr}
\renewcommand\refname{REFERENCES}
\bibliography{subgroup_generalizability_transportability}{}

%%%%%%%%%%%%%%%%%%%%%%%%%%%%%%%%%%%%%%%%%%%%%%%%%%%%%%%%%%%%
%%%%%%%%%%%%%%%%%%%%%%%%%%%TABLES%%%%%%%%%%%%%%%%%%%%%%%%%%%
%%%%%%%%%%%%%%%%%%%%%%%%%%%%%%%%%%%%%%%%%%%%%%%%%%%%%%%%%%%%
\clearpage
\section*{TABLES}

%$\delta_{\psi}(v,1,0)
\begin{table}[ht]
	\centering
	\caption{Subgroup-specific average treatment effects (risk differences), with corresponding 95\% confidence intervals, by history of myocardial infarction in the CASS target population of trial-eligible individuals (August 1975 to December 1996).}\label{table_cass_analyses1}
\begin{tabular}{@{}lcc@{}} 	\toprule
\textbf{Estimator} & \textbf{Previous MI} & \textbf{No previous MI} \\ \midrule
TRIAL  & -4.4\% (-12.1, 3.4) & 1.4\% (-6.2, 8.9) \\
OM   & -5.1\% (-13.1, 2.9) & 1.5\% (-6.0, 9.1)  \\
IPW1   & -4.6\% (-12.8, 3.5) & 1.0\% (-6.5, 8.6)   \\
IPW2  & -5.0\% (-13.0, 3.1) & 1.3\% (-6.3, 8.9)     \\
AIPW1  & -5.2\% (-13.3, 2.8) & 1.7\% (-5.9, 9.2) \\
AIPW2  & -5.2\% (-13.3, 2.8) & 1.7\% (-5.9, 9.2) \\ 
AIPW1 (GAM)  & -4.8\% (-12.9, 3.3) & 2.2\% (-5.5, 9.9) \\
AIPW2 (GAM)  & -4.8\% (-12.9, 3.3) & 2.2\% (-5.5, 9.9) \\ 
\bottomrule
\end{tabular}
	\caption*{We obtained 95\% Wald-style confidence intervals using standard errors estimated from 10,000 bootstrap samples. CASS = Coronary Artery Surgery Study; MI = myocardial infarction; OM = outcome-modeling estimator in equation \eqref{g_formula_estimator}; IPW1 = inverse probability weighted estimator in equation \eqref{IPW_estimator1}; IPW2 = inverse probability weighted estimator with normalized weights in equation \eqref{IPW_estimator2}; AIPW1 = augmented inverse probability weighted estimator in equation \eqref{DR_estimator1}; AIPW2 = augmented inverse probability weighted estimator with normalized weights in equation \eqref{DR_estimator2}. GAM = generalized additive models used to estimate the probability of trial participation and the expectation of the outcome, with a parametric model used to estimate the probability of treatment. For estimators in the table not labeled by GAM, we used parametric models.}
\end{table}

\begin{table}[ht]
	\centering
	\caption{Subgroup-specific average treatment effects (risk differences), with corresponding 95\% confidence intervals, by history of myocardial infarction in the non-randomized subset of the CASS target population of trial-eligible individuals (August 1975 to December 1996).}\label{table_cass_analyses2}
\begin{tabular}{@{}lcc@{}} \toprule
\textbf{Estimator} & \textbf{Previous MI} & \textbf{No previous MI} \\ \midrule
TRIAL & -4.4\% (-12.1, 3.4) & 1.4\% (-6.2, 8.9)  \\
OM     & -4.7\% (-12.8, 3.5) & 1.4\% (-6.4, 9.2) \\
IOW1  & -4.3\% (-12.7, 4.1) & 0.9\% (-6.9, 8.6)  \\
IOW2  & -4.9\% (-13.2, 3.4)  & 1.3\% (-6.5, 9.1) \\
AIOW1  & -5.0\% (-13.2, 3.3) & 1.6\% (-6.2, 9.4)  \\
AIOW2 & -5.0\% (-13.2, 3.3) & 1.6\% (-6.2, 9.4)   \\
AIOW1 (GAM)  & -4.2\% (-12.9, 4.5) &  2.3\% (-5.7, 10.4)  \\
AIOW2 (GAM)  & -4.2\% (-12.9, 4.5) & 2.3\% (-5.8, 10.4) \\ \bottomrule
\end{tabular}
	\caption*{We obtained 95\% Wald-style confidence intervals using standard errors estimated from 10,000 bootstrap samples. CASS = Coronary Artery Surgery Study; MI = myocardial infarction; OM = outcome-modeling estimator in equation \eqref{g_formula_estimator_S0}; IOW1 = inverse probability weighted estimator in equation \eqref{IPW_estimator1_S0}; IOW2 = inverse probability weighted estimator with normalized weights in equation \eqref{IPW_estimator2_S0}; AIOW1 = augmented inverse probability weighted estimator in equation \eqref{DR_estimator1_S0}; AIOW2 = augmented inverse probability weighted estimator with normalized weights in equation \eqref{DR_estimator2_S0}. GAM = generalized additive models used to estimate the probability of trial participation and the expectation of the outcome, with a parametric model used to estimate the probability of treatment. For estimators in the table not labeled by GAM, we used parametric models. }
\end{table}

%%%%%%%%%%%%%%%%%%%%%%%%%%%%%%%%%%%%%%%%%%%%%%%%%%%%%%%%%%%%%%%%%%%%%%%%%%%%%%
%\newpage
%\clearpage
%%%%%%%%%%%%%%%%%%%%%%%%%%%%%%%%%%%%%%%%%%%%%%%%%%%%%%%%%%%%%%%%%%%%%%%%%%%%%%

%\includepdf[pages={1,3,5}]{arxiv_submission_combined_appendix.pdf}

%%%%%%%%%%%%%%%%%%%%%%%%%%%%%%%%%%%%%%%%%%%%%%%%%%%%%%%%%%%%%%%%%%%%%%%%%%%%%%
% VERSIONING 
%%%%%%%%%%%%%%%%%%%%%%%%%%%%%%%%%%%%%%%%%%%%%%%%%%%%%%%%%%%%%%%%%%%%%%%%%%%%%%

\end{document}

\maketitle

\thispagestyle{empty}

\iffalse 
\clearpage
\setcounter{tocdepth}{1}
\tableofcontents
\fi

\thispagestyle{empty}

%%%%%%%%%%%%%%%%%%%%%%%%%%%%%%%%%%%%%%%%%%%%%%%%%%%%%%%%%%%%
%%%%%%%%%%%%%%%%%%%%%%%%%%APPENDIX%%%%%%%%%%%%%%%%%%%%%%%%%%
%%%%%%%%%%%%%%%%%%%%%%%%%%%%%%%%%%%%%%%%%%%%%%%%%%%%%%%%%%%%
\clearpage
\appendix 
\renewcommand{\thesection}{Web Appendix \arabic{section}}

\renewcommand{\thesubsection}{\arabic{section}.\arabic{subsection}}
\pagenumbering{arabic}% resets `page` counter to 1
%\renewcommand*{\thepage}{A\arabic{page}}
%%%%%%%%%%%%%%%%%%%%%%%%%%%%%%%%%%%%%%%%%%%%%%%%%%%%%%%%%%%%
%%%%%%%%%%%%%%%%%%%%%%%%%%%%%%%%%%%%%%%%%%%%%%%%%%%%%%%%%%%%
%%%%%%%%%%%%%%%%%%%%%%%%%%%%%%%%%%%%%%%%%%%%%%%%%%%%%%%%%%%%

\renewcommand{\theequation}{\arabic{section}.\arabic{equation}}

%%%%%%%%%%%%%%%%%%%%%%%%%%%%%%%%%%%%%%%%%%%%%%%%%%%%%%%%%%%%
%%%%%%%%%%%%%%%%%%%%%%%%%%%%%%%%%%%%%%%%%%%%%%%%%%%%%%%%%%%%

\clearpage

%%%%%%%%%%%%%%%%%%%%%%%%%%%%%%%%%%%%%%%%%%%%%%%%%%%%%%%%%%%%
\section{Data and estimation in non-nested trial study designs}\label{appendix:study_design}
%%%%%%%%%%%%%%%%%%%%%%%%%%%%%%%%%%%%%%%%%%%%%%%%%%%%%%%%%%%%

In a non-nested trial design, data from the trial participants and non-randomized individuals are collected separately. The data from the trial participants consist of $n_{\text{trial}}$ independent realizations of $(X, S=1, A, Y)$; the data from the sample of non-randomized individuals consist of $n_{\text{obs}}$ independent realizations of $(X, S=0)$.\footnote{In non-nested trial designs, the population of individuals with $S = 0$ is often referred to as \emph{the} target population and is not necessarily thought of as a subset of an overall target population.} Data from the trial participants and the non-randomized individuals are appended to form a composite dataset with sample size $n_{\text{trial}} + n_{\text{obs}}$. The data can be thought of as generated from a biased sampling model \cite{bickel1993efficient}, where randomized individuals and non-randomized individuals are sampled from an underlying population using possibly different, and unknown to the investigators, sampling fractions \cite{dahabreh2020transportingStatMed, dahabreh2021studydesigns}.

In the composite dataset, the estimators of $\phi(v,a)$ given in the main text, and this appendix, can, with minor modifications, be used to estimate subgroup-specific potential outcome means and average treatment effects in the population underlying the sample of non-randomized individuals. One necessary modification is that all conditional expectations and probabilities need to be estimated in the composite dataset; as such, they are estimated under the biased sampling model, mentioned above, and do not necessarily reflect the underlying population model. Another, is that all sums in the estimators are over the composite dataset, replacing $n$ with $n_{\text{trial}} + n_{\text{obs}}$.

%%%%%%%%%%%%%%%%%%%%%%%%%%%%%%%%%%%%%%%%%%%%%%%%%%%%%%%%%%%%
\clearpage
\section{Identification results}\label{appendix:identification}
%%%%%%%%%%%%%%%%%%%%%%%%%%%%%%%%%%%%%%%%%%%%%%%%%%%%%%%%%%%%

\subsection{Potential outcome means in the target population}
To derive the g-formula identification result, we re-write the subgroup-specific potential outcome mean for treatment $a$ using the observed data as follows:
\begin{equation*}
    \begin{split}
  \E[Y^a |  X \in v] %&= \E\big[\E[ Y^a | X,  X \in v] \big|  X \in v \big]   \\ 
         &= \E\big[\E[ Y^a | X] \big|  X \in v \big]   \\ 
         &= \E\big[\E[ Y^a | X, S=1] \big|  X \in v \big]   \\ 
        &= \E\big[\E[ Y^a | X, S=1,  A = a] \big|  X \in v \big]    \\ 
        &= \E\big[\E[Y | X, S=1,A = a] \big|  X \in v\big] \\
        &\equiv  \psi(v, a),
   \end{split}
\end{equation*}
where the first step follows from the law of total expectation, the second by mean exchangeability over $S$, the third by conditional mean exchangeability over $A$ in the trial, the last by consistency, and all quantities are well-defined because of the positivity conditions.

We will now show that $\psi(v,a) \equiv \E\big[ \E[Y | X, S=1, A = a] |  X \in v \big]$ has two inverse probability weighting re-expressions:
\begin{equation} \label{eq:ipw_re_expressions}
    \begin{split}
        \psi(v,a) &= \dfrac{1}{\Pr[X \in v]} \E\left[ \dfrac{ I(S = 1, A = a, X \in v)Y}{\Pr[S = 1| X] \Pr[A=a|X,S=1]}  \right] \\
        &= \Bigg\{  \E\left[\dfrac{I(S = 1, A = a, X \in v)}{\Pr[S = 1| X] \Pr[A=a|X,S=1]}\right] \Bigg\}^{-1} \E\left[ \dfrac{ I(S = 1, A = a, X \in v)Y}{\Pr[S = 1 | X] \Pr[A = a | X, S = 1]}  \right].
    \end{split}
\end{equation}
First, we note that under the positivity conditions, 
\begin{equation*}
    \begin{split}
        \psi(v,a) &= \E\big[\E[Y | X, S=1, A = a] | X \in v\big]  \\
            &=\dfrac{1}{\Pr[X \in v]} \E\Big[I(X \in v)  \E[Y|X, S=1, A=a] \Big] \\
            &=\dfrac{1}{\Pr[X \in v]} \E\Bigg[ I(X \in v) \E\left[ \dfrac{ I(S = 1, A=a)Y}{\Pr[S = 1, A=a | X]} \Big| X  \right]  \Bigg]  \\
            &=\dfrac{1}{\Pr[X \in v]} \E\Bigg[ \E\left[ \dfrac{ I(S = 1, A=a,X \in v)Y}{\Pr[S = 1, A=a | X]} \Big| X  \right]  \Bigg]  \\
            &= \dfrac{1}{\Pr[X \in v]} \E\left[ \dfrac{ I(S = 1, A = a, X \in v)Y}{\Pr[S = 1 | X] \Pr[A = a | X, S = 1]}  \right],
    \end{split}
\end{equation*}
which establishes the first equality in \eqref{eq:ipw_re_expressions}. 

Next, we note that 
\begin{equation*}
    \begin{split}
        \E\left[\dfrac{I(S = 1, A = a,X \in v)}{\Pr[S = 1 | X] \Pr[A = a | X, S = 1]}\right] &=  \E\Bigg[ \E\left[\dfrac{I(S = 1, A = a,X \in v)}{\Pr[S = 1 | X] \Pr[A = a | X, S = 1]}  \Big| X \right] \Bigg] \\
         &=  \E\Bigg[ \dfrac{I(X \in v)}{\Pr[S = 1 | X] \Pr[A = a | X, S = 1]} \E[I(S=1, A = a) | X ] \Bigg] \\
          &= \Pr[X \in v].
    \end{split}
\end{equation*}
Thus, we can also write
\begin{equation*}
    \psi(v,a) =  \Bigg\{  \E\left[\dfrac{I(S = 1, A = a, X \in v)}{\Pr[S = 1 | X] \Pr[A = a | X , S = 1]}\right] \Bigg\}^{-1} \E\left[ \dfrac{ I(S = 1, A = a, X \in v)Y}{\Pr[S = 1 | X] \Pr[A = a | X, S = 1]}  \right],
\end{equation*}
which establishes the second equality in \eqref{eq:ipw_re_expressions}.

\subsection{Potential outcome means in the non-randomized subset}
To derive the g-formula identification result, we re-write the subgroup-specific potential outcome mean for treatment $a$ using the observed data as follows:
\begin{equation*}
    \begin{split}
  \E[Y^a |  X \in v, S=0] % &= \E\big[\E[ Y^a | X,  X \in v,S=0] \big|  X \in v, S=0 \big]   \\ 
         &= \E\big[\E[ Y^a | X,S=0] \big|  X \in v,S=0 \big]   \\ 
         &= \E\big[\E[ Y^a | X, S=1] \big|  X \in v,S=0 \big]   \\ 
        &= \E\big[\E[ Y^a | X, S=1,  A = a] \big|  X \in v,S=0 \big]    \\ 
        &= \E\big[\E[Y | X, S=1,A = a] \big|  X \in v, S=0\big] \\
        &\equiv  \phi(v, a),
   \end{split}
\end{equation*}
where the first step follows from the law of total expectation, the second by mean exchangeability over $S$, the third by conditional mean exchangeability over $A$ in the trial, the last by consistency, and all quantities are well-defined because of the positivity condition.

We will now show that $\phi(v,a) \equiv \E\big[ \E[Y | X, S=1, A = a] |  X \in v, S=0 \big]$ has two inverse probability weighting re-expressions:
\begin{equation} \label{eq:ipw_re_expressions_S0}
    \begin{split}
        \phi(v,a) &= \dfrac{1}{\Pr[X \in v,S=0]} \E\left[ \dfrac{ I(S = 1, A = a, X \in v) \Pr[S=0|X]Y}{\Pr[S = 1 | X] \Pr[A = a | X, S = 1]}  \right] \\
        &= \Bigg\{  \E\left[\dfrac{I(S = 1, A = a, X \in v)\Pr[S=0|X]}{\Pr[S = 1 | X] \Pr[A = a | X, S = 1]}\right] \Bigg\}^{-1} \E\left[ \dfrac{ I(S = 1, A = a, X \in v) \Pr[S=0|X]Y}{\Pr[S = 1 | X] \Pr[A = a | X, S = 1]}  \right].
    \end{split}
\end{equation}

First, we note that 
\begin{equation*}
    \begin{split}
        \phi(v,a) &= \E\big[\E[Y | X, S=1, A = a] | X \in v, S=0\big]  \\
            &=\dfrac{1}{\Pr[X \in v, S=0]} \E\Big[I(X \in v,S=0)  \E[Y|X, S=1, A=a] \Big] \\
            &=\dfrac{1}{\Pr[X \in v,S=0]} \E\Bigg[ I(X \in v,S=0) \E\left[ \dfrac{ I(S = 1, A=a)Y}{\Pr[S = 1 | X] \Pr[A = a | X, S = 1]} \Big| X  \right]  \Bigg]  \\
            &=\dfrac{1}{\Pr[X \in v,S=0]} \E\Bigg[I(S=0)  \E\left[\dfrac{ I(S = 1, A=a,X \in v)Y}{\Pr[S = 1 | X] \Pr[A = a | X, S = 1]} \Big| X  \right]  \Bigg]  \\
            &=\dfrac{1}{\Pr[X \in v,S=0]} \E\Bigg[\Pr[S=0|X]  \E\left[\dfrac{ I(S = 1, A=a,X \in v)Y}{\Pr[S = 1 | X] \Pr[A = a | X, S = 1]} \Big| X  \right]  \Bigg]  \\
            &=\dfrac{1}{\Pr[X \in v,S=0]} \E\Bigg[ \E\left[\dfrac{ I(S = 1, A=a,X \in v) \Pr[S=0|X] Y}{\Pr[S = 1 | X] \Pr[A = a | X, S = 1]} \Big| X  \right]  \Bigg]  \\
            &= \dfrac{1}{\Pr[X \in v,S=0]} \E\left[ \dfrac{ I(S = 1, A = a, X \in v)\Pr[S=0|X] Y}{\Pr[S = 1 | X] \Pr[A = a | X, S = 1]}  \right],
    \end{split}
\end{equation*}
which establishes the first equality in \eqref{eq:ipw_re_expressions}. Next, we note that 
\begin{equation*}
    \begin{split}
        \E\left[\dfrac{I(S = 1, A = a,X \in v)\Pr[S=0|X] }{\Pr[S = 1 | X] \Pr[A = a | X , S = 1]}\right] &=  \E\Bigg[ \E\left[\dfrac{I(S = 1, A = a,X \in v)\Pr[S=0|X] }{\Pr[S = 1 | X] \Pr[A = a | X , S = 1]}  \Big| X \right] \Bigg] \\
         &=  \E\Bigg[ \dfrac{I(X \in v) \Pr[S = 0 | X]}{\Pr[S = 1 | X] \Pr[A = a | X, S = 1]} \E[I(S=1, A = a) | X ] \Bigg] \\
          &=  \E\big[ I(X \in v) \Pr[S = 0 | X] \big]  \\
           &=  \E\big[ \E[ I(X \in v , S = 0) | X] \big]  \\
          &= \Pr[X \in v,S=0].
    \end{split}
\end{equation*}
Thus, we can also write
\begin{equation*}
    \phi(v,a) =  \Bigg\{  \E\left[\dfrac{I(S = 1, A = a, X \in v)\Pr[S = 0 | X]}{\Pr[S = 1 | X] \Pr[A = a | X , S = 1]}\right] \Bigg\}^{-1} \E\left[ \dfrac{ I(S = 1, A = a, X \in v)Y \Pr[S = 0| X]}{\Pr[S = 1 | X] \Pr[A = a | X, S = 1]}  \right],
\end{equation*}
which establishes the second equality in \eqref{eq:ipw_re_expressions}.

%%%%%%%%%%%%%%%%%%%%%%%%%%%%%%%%%%%%%%%%%%%%%%%%%%%%%%%%%%%%
\clearpage
\section{Weighted estimators using saturated regression models}\label{appendix:saturated_IPW_estimators}
\setcounter{figure}{0}
\setcounter{equation}{0}
\renewcommand{\figurename}{Web Figure}
\renewcommand{\tablename}{Web Table}
%%%%%%%%%%%%%%%%%%%%%%%%%%%%%%%%%%%%%%%%%%%%%%%%%%%%%%%%%%%%

\subsection{Potential outcome means and treatment effects in the target population}

In the main text we provided the following equation for the inverse probability weighting estimator with normalized weights: \cite{hajek1971comment}:
\begin{equation*} \label{IPW_estimator2}
  \widehat {\psi}_{\text{\tiny{IPW2}}}(v,a) = \left\{ \sum_{i=1}^{n} \widehat w_{v,a}(X_i, S_i, A_i)  \right\}^{-1} \sum\limits_{i = 1}^{n} \widehat w_{v,a}(X_i, S_i, A_i)  Y_i.
\end{equation*}

An estimator equivalent to $\widehat {\psi}_{\text{\tiny{IPW2}}}(v,a)$ can be obtained by fitting an appropriate saturated mean regression model for the outcome $Y$ conditional on indicators (dummy variables) for treatment $A$ and subgroup membership (for each subgroup $v \in \{v_1, \ldots, v_k\}$) and weights equal to the inverse of the product of the probability of trial participation times the probability of receiving the treatment actually received among trial participants for observations in the trial, and 0 otherwise \cite{robins2000a}. A saturated mean model should include main effects for treatment and the subgroup indicators, as well as all possible interactions between them. The regression can be estimated just among trial participants because non-randomized individuals receive weight zero.

As an example, consider a binary treatment $A$; two subgroups $v_1$ and $v_2$, such that $\mathcal X = v_1 \cup v_2$, and a continuous outcome $Y$. We estimate the following saturated linear regression model among trial participants, $$\E[Y |  X, A , S = 1] = \alpha_0 +  \alpha_1 A +  \alpha_2 I(X \in v_1) +  \alpha_3 A \times I(X \in v_1),$$ using estimated observation-level weights equal to $$ \dfrac{S_i}{\widehat p(X_i)} \times \left\{ \dfrac{A_i}{\widehat e_1(X_i)} + \dfrac{1 - A_i}{\widehat e_0(X_i)} \right\}.$$ To connect these weights with the notation in the main text, note that $$ \dfrac{S_i}{\widehat p(X_i)} \times \left\{ \dfrac{A_i}{\widehat e_1(X_i)} + \dfrac{1 - A_i}{\widehat e_0(X_i)} \right\} = \sum\limits_{v,a} \widehat w_{v,a}(X_i, S_i, A_i).$$ The estimated regression coefficients from this model, $\widehat \alpha_0, \ldots, \widehat \alpha_3$, can be used to obtain the estimators $\widehat \psi_{\text{\tiny IPW2}}(v,a)$, for $a=0,1$ and $v=0,1$, as follows: 
\begin{equation*}
    \begin{split}
        \widehat \psi_{\text{\tiny IPW2}}(0,0) &= \widehat \alpha_0, \\
        \widehat \psi_{\text{\tiny IPW2}}(0,1) &= \widehat \alpha_0 + \widehat \alpha_1, \\
        \widehat \psi_{\text{\tiny IPW2}}(1,0) &= \widehat \alpha_0 + \widehat \alpha_2, \\
        \widehat  \psi_{\text{\tiny IPW2}}(1,1) &= \widehat \alpha_0 + \widehat \alpha_1 + \widehat \alpha_2 + \widehat \alpha_3. 
    \end{split}
\end{equation*}
Furthermore, the estimated coefficients can be used to estimate subgroup-specific average treatment effects. For example, the average treatment effect in the subgroup with $v=0$ can be estimated by $\widehat \psi_{\text{\tiny IPW2}}(0,1) - \widehat \psi_{\text{\tiny IPW2}}(0,0) = \widehat \alpha_1$. Similarly, the average treatment effect in the subgroup with $v = 1$, can be estimated by
$\widehat \psi_{\text{\tiny IPW2}}(1,1) - \widehat \psi_{\text{\tiny IPW2}}(1,0)   = \widehat \alpha_1 + \widehat \alpha_3.$ 

To properly account for the estimation of the weights, inference should be based on resampling methods (e.g., the non-parametric bootstrap \cite{efron1994introduction}, estimating the weights and the saturated model in each bootstrap iteration \cite{robins2000a, lunceford2004}). An alternative is to jointly estimate the weights and the potential outcome means/average treatment effects (e.g., using standard stacked M-estimation methods \cite{stefanski2002, lunceford2004}). Using the robust (Huber-White) variance estimator \cite{huber1967behavior, white1996estimation} when fitting the weighted regression model with estimated weights does not explicitly account for the estimation of the weights but is expected to produce conservative estimates of uncertainty \cite{robins2000a}.

\subsection{Potential outcome means and treatment effects in the non-randomized subset of the target population\footnote{The approach we describe in this subsection is similar to the one in reference \cite{mehrotra2021transporting}, but we use weights that involve the inverse of the estimated probability of receiving the treatment actually received (among trial participants) to improve efficiency.}}

In the main text, we provided the following equation for the inverse odds weighting estimator with normalized weights: 
\begin{equation*} \label{IPW_estimator2_S0}
  \widehat {\phi}_{\text{\tiny{IOW2}}}(v,a) = \left\{ \sum_{i=1}^{n}  \widehat o_{v,a}(X_i, S_i,A_i) \right\}^{-1} \sum\limits_{i = 1}^{n}  \widehat o_{v,a}(X_i, S_i,A_i) Y_i.
\end{equation*}

Similar to the previous section of this Appendix, an estimator equivalent to $\widehat {\phi}_{\text{\tiny{IOW2}}}(v,a)$ can be obtained by fitting an appropriate saturated mean regression model for the outcome $Y$ conditional on indicators (dummy variables) for treatment $A$ and subgroup membership (for each subgroup $v \in \{v_1, \ldots, v_k\}$) and weights equal to the inverse of the product of the odds of trial participation times the probability of receiving the treatment actually received among trial participants for observations in the trial, and 0 otherwise \cite{robins2000a}. As above, a saturated mean model should include main effects for treatment and the subgroup indicators, as well as all possible interactions between them. The model can be estimated just among trial participants because non-randomized individuals receive weight zero. As an example, consider a binary treatment $A$; two subgroups $v_1$ and $v_2$, such that $\mathcal X = v_1 \cup v_2$, and a continuous outcome $Y$. We estimate the following saturated linear regression model among trial participants, $$\E[Y |  X, A , S = 1] = \alpha_0 +  \alpha_1 A +  \alpha_2 I(X \in v_1) +  \alpha_3 A \times I(X \in v_1),$$ using estimated observation-level weights equal to $$ \dfrac{S_i \times \{ 1-  \widehat p(X_i) \} }{\widehat p(X_i)} \times \left\{ \dfrac{A_i}{\widehat e_1(X_i)} + \dfrac{1 - A_i}{\widehat e_0(X_i)} \right\}.$$ To connect these weights with the notation in the main text, note that $$ \dfrac{S_i \times \{ 1-\widehat p(X_i) \} }{\widehat p(X_i)} \times \left\{ \dfrac{A_i}{\widehat e_1(X_i)} + \dfrac{1 - A_i}{\widehat e_0(X_i)} \right\} = \sum\limits_{v,a} \widehat o_{v,a}(X_i, S_i, A_i).$$ The estimated regression coefficients from this model, $\widetilde \alpha_0, \ldots, \widetilde \alpha_3$, can be used to obtain the estimators $\widehat \psi_{\text{\tiny IOW2}}(v,a)$, for $a=0,1$ and $v=0,1$, as follows: 
\begin{equation*}
    \begin{split}
        \widehat \phi_{\text{\tiny IOW2}}(0,0) &= \widetilde \alpha_0, \\
        \widehat \phi_{\text{\tiny IOW2}}(0,1) &= \widetilde \alpha_0 + \widetilde \alpha_1, \\
        \widehat \phi_{\text{\tiny IOW2}}(1,0) &= \widetilde \alpha_0 + \widetilde \alpha_2, \\
        \widehat  \phi_{\text{\tiny IOW2}}(1,1) &= \widetilde \alpha_0 + \widetilde \alpha_1 + \widetilde \alpha_2 + \widetilde \alpha_3. 
    \end{split}
\end{equation*}
Furthermore, these estimated coefficients can be used to estimate subgroup-specific average treatment effects. For example, the average treatment effect in the subgroup with $v=0$ can be estimated by $\widehat \phi_{\text{\tiny IPW2}}(0,1) - \widehat \phi_{\text{\tiny IPW2}}(0,0) = \widetilde \alpha_1$. Similarly, the average treatment effect in the subgroup with $v = 1$, can be estimated by
$\widehat \phi_{\text{\tiny IPW2}}(1,1) - \widehat \phi_{\text{\tiny IPW2}}(1,0)   = \widetilde \alpha_1 + \widetilde \alpha_3.$ 
Inference can be carried out as described in the previous subsection of this appendix.

%%%%%%%%%%%%%%%%%%%%%%%%%%%%%%%%%%%%%%%%%%%%%%%%%%%%%%%%%%%%
\clearpage
\section{Weighted multivariable regression estimators}\label{appendix:additional_estimators}
\setcounter{figure}{0}
\setcounter{equation}{0}
\renewcommand{\figurename}{Web Figure}
\renewcommand{\tablename}{Web Table}
%%%%%%%%%%%%%%%%%%%%%%%%%%%%%%%%%%%%%%%%%%%%%%%%%%%%%%%%%%%%

\subsection{Potential outcome means in the target population}

A third doubly robust estimator relies on fitting a multivariable regression model for the outcome estimated using inverse probability weighting, followed by standardization over the distribution of baseline covariates \cite{robins2007, wooldridge2007}. The potential outcome mean is estimated as 
\begin{equation} \label{DR_estimator3}
  \widehat {\psi}_{\text{\tiny{AIPW3}}}(v,a) = \left\{ \sum_{i=1}^{n} I(X_i \in v) \right\}^{-1}  \sum\limits_{i = 1}^{n} I(X_i \in v) g_{a}(X_i; \widehat\theta_a),
\end{equation}
where $g_{a}(X; \widehat \theta_a)$ is an estimator for $\E[Y | X, S=1, A = a]$, with a vector of estimated model parameters $\widehat\theta_a$, from an inverse probability weighted outcome regression, with weights $$\dfrac{S_i}{\widehat p(X_i)} \times \left\{ \dfrac{A_i}{\widehat e_1(X_i)} + \dfrac{1 - A_i}{\widehat e_0(X_i)} \right\}.$$ When the outcome is modeled in the linear exponential family with a canonical link, and estimation is by quasi-likelihood methods (e.g., linear or logistic regression) \cite{gourieroux1984}, this estimator has the double robustness property.

\subsection{Potential outcome means in the non-randomized subset of the target population}

A third doubly robust estimator relies on fitting a multivariable regression model for the outcome estimated using inverse odds weighting, followed by standardization over the distribution of baseline covariates \cite{robins2007, wooldridge2007}. The potential outcome mean is estimated as 
\begin{equation} \label{DR_estimator3_S0}
  \widehat {\phi}_{\text{\tiny{AIOW3}}}(v,a) = \left\{ \sum_{i=1}^{n} I(X_i \in v, S = 0) \right\}^{-1}  \sum\limits_{i = 1}^{n} I(X_i \in v, S = 0) g_{a}(X_i; \widetilde\theta_a),
\end{equation}
where $g_{a}(X; \widetilde \theta_a)$ is an estimator for $\E[Y | X, S=1, A = a]$, with a vector of estimated model parameters $\widetilde\theta_a$, from an inverse odds weighted outcome regression, with weights $$\dfrac{S_i \times \{ 1-\widehat p(X_i) \} }{\widehat p(X_i)} \times \left\{ \dfrac{A_i}{\widehat e_1(X_i)} + \dfrac{1 - A_i}{\widehat e_0(X_i)} \right\}.$$ When the outcome is modeled in the linear exponential family with a canonical link, and estimation is by quasi-likelihood methods (e.g., linear or logistic regression) \cite{gourieroux1984}, this estimator has the double robustness property.

%%%%%%%%%%%%%%%%%%%%%%%%%%%%%%%%%%%%%%%%%%%%%%%%%%%%%%%%%%%%
\clearpage
\setcounter{figure}{0}
\setcounter{equation}{0}
\section{Additional results from the CASS analyses}\label{appendix:cass_baselines}
%%%%%%%%%%%%%%%%%%%%%%%%%%%%%%%%%%%%%%%%%%%%%%%%%%%%%%%%%%%%

\subsection*{Baseline covariates}
\begin{table}[H]
\caption{CASS baseline table (August 1975 to December 1996). $S=1 $ indicates randomized ($S=0$ indicates non-randomized) $A=1$ indicates surgical therapy ($A=0$ indicates medical therapy).}
\label{cass_baseline}
\centering
\begin{tabular}{@{}lcccc@{}}
\toprule
 & $S=1, A=1$ & $S=1, A=0$ & $S=1$ & $S=0$ \\ \midrule
Number of patients  & 368 & 363 & 731 & 955 \\
Age & 51.42 (7.24) & 50.92 (7.41) & 51.17 (7.32) & 50.89 (7.73) \\
History of angina & 285 (77.4) & 282 (77.7) & 567 (77.6) & 760 (79.6) \\
%Taken beta-blocker regularly & 163 (44.3) & 152 (41.9) & 315 (43.1) & 508 (53.2) \\
%Taken diuretic regularly & 63 (17.1) & 50 (13.8) & 113 (15.5) & 145 (15.2) \\
Ejection fraction & 60.86 (13.04) & 59.83 (12.78) & 60.35 (12.91) & 60.16 (12.25) \\
%Employed full-time & 264 (71.7) & 233 (64.2) & 497 (68.0) & 632 (66.2) \\
%High physical labor job & 151 (41.0) & 142 (39.1) & 293 (40.1) & 340 (35.6) \\
%Low mental labor job & 129 (35.1) & 135 (37.2) & 264 (36.1) & 320 (33.5) \\
%High mental labor job & 88 (23.9) & 86 (23.7) & 174 (23.8) & 295 (30.9) \\
%Left ventricular wall score & 7.44 (2.89) & 7.30 (2.78) & 7.37 (2.84) & 7.07 (2.69) \\
%Taken nitrates regularly & 205 (55.7) & 196 (54.0) & 401 (54.9) & 528 (55.3) \\
History of previous MI & 209 (56.8) & 228 (62.8) & 437 (59.8) & 549 (57.5) \\
%Male & 35 (9.5) & 37 (10.2) & 72 (9.8) & 87 (9.1) \\
%Never smoked & 62 (16.8) & 54 (14.9) & 116 (15.9) & 157 (16.4) \\
%Formerly smoked & 164 (44.6) & 157 (43.3) & 321 (43.9) & 451 (47.2) \\
%Presently smokes & 142 (38.6) & 152 (41.9) & 294 (40.2) & 347 (36.3) \\
%High limitation of activities & 165 (44.8) & 173 (47.7) & 338 (46.2) & 441 (46.2) \\
%High recreational activity & 228 (62.0) & 219 (60.3) & 447 (61.1) & 616 (64.5) \\
%Confirmed hypertension & 118 (32.1) & 108 (29.8) & 226 (30.9) & 260 (27.2) \\
%No diabetes & 325 (88.3) & 328 (90.4) & 653 (89.3) & 873 (91.4) \\
%Uncertain diabetes & 13 (3.5) & 7 (1.9) & 20 (2.7) & 23 (2.4) \\
%Confimed diabetes & 30 (8.2) & 28 (7.7) & 58 (7.9) & 59 (6.2) \\
LMCA percent obstruction & 4.27 (11.87) & 2.78 (9.55) & 3.53 (10.80) & 5.76 (14.50) \\
PLMA percent obstruction & 36.44 (38.04) & 34.89 (36.95) & 35.67 (37.49) & 39.14 (38.73) \\
Any diseased proximal vessels & 222 (60.3) & 230 (63.4) & 452 (61.8) & 608 (63.7) \\ \bottomrule
%Systolic blood pressure & 130.28 (17.40) & 130.34 (18.72) & 130.31 (18.06) & 129.80 (18.23) \\ \bottomrule
\end{tabular}
\caption*{MI= myocardial infarction; LMCA = left main coronary artery; PLMA= proximal left anterior artery. For continuous variables we report the mean (standard deviation); for binary variables we report the number of individuals (percentage).}
\end{table}

\clearpage
\subsection*{Estimates of potential outcome means}

We report estimates of the subgroup-specific mortality risk for each treatment group, for the entire target population and its non-randomized subset. All equation numbers in the captions refer to the main text.

\vspace{0.3in}

\begin{table}[ht!]
	\footnotesize
		\centering
		\caption{Potential outcome mean estimates for subgroup analyses in the CASS study (August 1975 to December 1996), for previous myocardial infarction.}
		\label{appendix_table_cass_analyses1}
\begin{tabular}{@{}llllll@{}}
\toprule
\multicolumn{1}{l}{Estimator} & \multicolumn{2}{c}{$v=1$} & \multicolumn{2}{c}{$v=0$} &  \\ 
 & \multicolumn{1}{c}{$a=1$} & \multicolumn{1}{c}{$a=0$} & \multicolumn{1}{c}{$a=1$} & \multicolumn{1}{c}{$a=0$} &  \\ \midrule
\multicolumn{6}{l}{Trial-only population} \\ \midrule
TRIAL  & 20.6\%  (15.2, 26.1) & 25.0\% (19.5, 30.7) & 13.2\% (8.2, 18.6) & 11.9\% (6.6, 17.6) \\ \midrule
\multicolumn{6}{l}{Entire target population} \\ \midrule
OM  & 20.6\% (14.9, 26.3) & 25.7\% (20.1, 31.3) & 13.6\% (8.2, 19.0) & 12.1\% (6.7, 17.5) &  \\
IPW1 & 21.1\% (15.2, 26.9) & 25.7\% (20.1, 31.4) & 13.1\% (7.8, 18.3) & 12.0\% (6.5, 17.6) &  \\
IPW2 & 20.8\% (15.1, 26.5) & 25.8\% (20.1, 31.4) & 13.3\% (8.0, 18.6) & 12.0\% (6.5, 17.5)  &  \\
AIPW1  & 20.6\% (14.9, 26.3) & 25.8\% (20.2, 31.4) & 13.6\% (8.1, 19.0) & 11.9\% (6.6, 17.2) &  \\
AIPW2  & 20.6\% (14.9, 26.3) & 25.8\% (20.2, 31.4) & 13.6\% (8.1, 19.0) & 11.9\% (6.6, 17.2) &  \\
AIPW1 (GAM)  & 21.1\% (15.3, 26.9) & 25.8\% (20.2, 31.5) & 13.9\% (8.3, 19.6) & 11.7\% (6.4, 17.0) &  \\
AIPW2 (GAM)  & 21.1\% (15.3, 26.9) & 25.8\% (20.2, 31.5) & 13.9\% (8.3, 19.6) & 11.7\% (6.4, 17.0) &  \\ \midrule
\multicolumn{6}{l}{Non-randomized subset of the target population} \\ \midrule
OM  & 20.8\% (14.9, 26.7) & 25.5\% (19.7, 31.2) & 13.9\% (8.3, 19.4) & 12.4\% (6.9, 18.0) &  \\
IOW1 & 21.4\% (15.3, 27.4) & 25.7\% (19.9, 31.5) & 13.1\% (7.8, 18.4) & 12.3\% (6.6, 17.9) &  \\
IOW2  & 20.9\% (15.0, 26.8) & 25.8\% (20.0, 31.6) & 13.5\% (8.0, 18.9) & 12.2\% (6.6, 17.8) &  \\
AIOW1 & 20.6\% (14.7, 26.6) & 25.6\% (19.8, 31.4) & 13.9\% (8.2, 19.6) & 12.2\% (6.8, 17.7)  &  \\
AIOW2 & 20.6\% (14.7, 26.6) & 25.6\% (19.8, 31.4) & 13.9\% (8.2, 19.6) & 12.2\% (6.8, 17.7)  &  \\
AIOW1 (GAM)  & 21.4\% (15.0, 27.8) & 25.6\% (19.5, 31.7) & 14.3\% (8.3, 20.3) & 12.0\% (6.4, 17.5) &  \\
AIOW2 (GAM)  & 21.4\% (15.0, 27.8) & 25.6\% (19.5, 31.7) & 14.3\% (8.3, 20.3) & 12.0\% (6.4, 17.5) &  \\ \bottomrule
\end{tabular}
	\caption*{We obtained 95\% Wald-style confidence intervals using standard errors estimated from 10,000 bootstrap samples. CASS = Coronary Artery Surgery Study; $v=1$ indicates history of myocardial infarction ($v=0$ indicates no history of myocardial infarction); GAM = generalized additive models used to estimate the probability of trial participation and the expectation of the outcome, with a parametric model used to estimate the probability of treatment. For estimators in the table not labeled by GAM, we used parametric models. Here $a=1$ is surgical therapy; $a=0$ is medical therapy. See main text for descriptions of estimators.}
\end{table}

%%%%%%%%%%%%%%%%%%%%%%%%%%%%%%%%%%%%%%%%%%%%%%%%%%%%%%%%%%%%
\clearpage
\setcounter{figure}{0}
\setcounter{equation}{0}
\section{Code and data}\label{appendix:code}
%%%%%%%%%%%%%%%%%%%%%%%%%%%%%%%%%%%%%%%%%%%%%%%%%%%%%%%%%%%%

\paragraph{Code for empirical analyses:} We provide \texttt{R} \cite{currentRcitation} code to implement all the estimators described in the paper for the $\texttt{R}$ environment on GitHub: [link removed for peer review]. 
%We provide \texttt{R} code to reproduce the analyses in Section 7 of the paper. Specifically, we provide the following files:
%https://github.com/serobertson/ExtendingSubgroupEffects

%\iffalse 

\begin{itemize}
\item[]   \texttt{00\_subgroup\_generalizability\_source\_code\_tablecenters.R} contains the source code for all the estimators. 

\item[]  \texttt{01\_subgroup\_transportability.R} runs the source code, using a simulated dataset for illustration.
\end{itemize}

%\fi

\paragraph{Data availability:} The CASS study data are not publicly available, but they can be obtained from the National Heart, Lung, and Blood Institute (NHLBI) Biologic Specimen and Data Repository Information Coordinating Center (\url{https://biolincc.nhlbi.nih.gov/studies/cass}; last accessed April 30, 2021).

%%%%%%%%%%%%%%%%%%%%%%%%%%%%%%%%%%%%%%%%%%%%%%%%%%%%%%%%%%%%
%%%%%%%%%%%%%%%%%%%%%%%%%REFERENCES%%%%%%%%%%%%%%%%%%%%%%%%%
%%%%%%%%%%%%%%%%%%%%%%%%%%%%%%%%%%%%%%%%%%%%%%%%%%%%%%%%%%%%
\clearpage
\bibliographystyle{unsrt}
\bibliography{subgroup_generalizability_transportability}{}
%%%%%%%%%%%%%%%%%%%%%%%%%%%%%%%%%%%%%%%%%%%%%%%%%%%%%%%%%%%%